\documentclass[onecolumn,aps,floatfix,pra,superscriptaddress,nofootinbib]{revtex4}

\usepackage{color}
\usepackage{graphicx}
\usepackage{bm}
\usepackage{amsmath}
\usepackage{enumerate}

\newcommand{\cl}{ \text{cl} }
\newcommand{\ti}{ \tilde }
\newcommand{\ep}{ \epsilon }

\begin{document}

\title{Quantum-classical transition in dissipative systems through scaled trajectories}
\author{S. V. Mousavi}
\email{vmousavi@qom.ac.ir}
\affiliation{Department of Physics, The University of Qom, P. O. Box 37165, Qom, Iran}
\author{S. Miret-Art\'es}
\email{s.miret@iff.csic.es}
\affiliation{Instituto de F\'isica Fundamental, Consejo Superior de
Investigaciones Cient\'ificas, Serrano 123, 28006 Madrid, Spain}

\begin{abstract}

A nonlinear quantum-classical transition wave equation is proposed for dissipative systems within the Caldirola-Kanai model. Equivalence of this transition equation to a scaled Schr\"{o}dinger equation is proved. The dissipative dynamics is then studied in terms of what we call scaled trajectories following the standard procedure used in Bohmian mechanics. These trajectories depend on a continuous parameter allowing us a smooth transition from Bohmian to classical trajectories. Arrival times and actual momentum distribution functions are also analyzed. The propagation of a Gaussian wave packet in a viscid medium under the presence of constant, linear and harmonic potentials is studied. The gradual decoherence process and localization are easily visualized and understood within this theoretical framework.

\end{abstract}
\maketitle


\section{Introduction}

Decoherence and emergence of the classical behavior in open quantum systems is a very broad 
and active field of research \cite{percival,weiss,breuer,Razavy-book-2005,salva1,salva2,salva3}. 
Three main different approaches to deal with quantum dissipative/stochastic dynamics are 
considered in literature \cite{salva3}. First, the system-plus-environment model where the
physical system of interest and its surrounding is considered as a isolated system, 
both parts being in continuous interaction. Open system dynamics are described following one of the 
five standard pictures of quantum mechanics: Schr\"odinger, Heisenberg, Dirac (interaction), 
Feynman and de Broglie-Bohm. 
In the Scr\"odinger and Dirac pictures, master equations for the reduced system where the
bath degrees of freedom have been traced out are obtained. A Markovian behavior of the 
corresponding equations  is reached when the bath is assumed to have no memory, as displayed 
by the Linblad equation. Non-Markovian equations lead to a loss of analytical expressions
as well as larger computational times. Following the path integral formulation, the time evolution
of the density matrix is also used with great success.  In the Heisenberg picture, the quantum
(generalized or standard) Langevin equation framework has been widely developed. 
Nowadays, the application of these stochastic quantum formalisms
are being addressed to more and more complex systems. 
If the wave function is seen as a stochastic process in Hilbert space, 
the dynamics is no longer described by a master equation but by the
It$\hat{o}$ stochastic differential equation and the realizations of the
underlying stochastic processes are called quantum trajectories. Unfortunately, 
this is the same denomination used in Bohmian mechanics and can lead to confussion
\cite{Holland-book-1993,salva1,salva2}.
The Bohmian picture allows us to also describe the corresponding dynamics in terms of trajectories,
enabling a quite straightforward correspondence with classical trajectories. 
Second, nonlinear wave mechanics which constitutes one of the most investigated topics 
in quantum mechanics \cite{salva3}. And third, effective time dependent
Hamiltonians. Dissipation is described by explicitly time dependent Lagrangians and/or 
Hamiltonians, thus avoiding to deal with the environment degrees of freedom. This approach
allows one to preserve the canonical formalism which can be a good starting point to reach
the quantum analog of the corresponding dissipative dynamics. The so-called Caldirola-Kanai (CK)
model \cite{caldirola,kanai} is considered a Hamiltonin formulation of the Langevin 
equation with zero fluctuations. In this framework, the system undergoes a gradual decay until 
its total energy is completely and irreversibly lost by dissipation. The dynamics is then  stopped 
and localized. This model can provide us a nice illustration of the 
dynamics of a particle on quantum viscid media \cite{SaCaRoLoMi-AP-2014}.

In this work, the CK model is considered to analyze the quantum-classical transition in dissipative
systems. For this goal, a continuous and gradual transition is proposed through trajectories within
the Bohmian framework. Our starting point is the so-called classical Schr\"odinger equation  
\cite{Ro-AJP-1964}. Rozen  argued that, in the large mass limit, the Schr\"{o}dinger equation
should be replaced by another one, known as classical wave equation,
which is equivalent to the continuity and classical Hamilton-Jacobi equations. This classical 
wave equation contains a non-linear term in the wave function which in general prevents 
superposition of different states.
Following Rozen, these authors  \cite{MuMi-PS-2015} have studied quantum-classical
transition in terms of arrival times in the scattering of a non-Gaussian wave packet by a rectangular
barrier.
Recently, a generalization of Rozen's work has been carried out.
In this regard, a wave equation known as transition equation which can be tuned
to describe both classical and quantum behavior in a continuous way has been proposed 
\cite{RiSchMaVaBa-PRA-2014}, and quantum-like behaviour of the corresponding
equation has been shown to be  equivalent to a linear Schr\"{o}dinger
equation with a rescaled Planck's constant \cite{RiSchMaVaBa-PRA-2014}.
Chou \cite{Ch-AP-2016} has studied this quantum-classical transition 
for wave packet interference within the Bohmian framework.
All of this work has been carried out for conservative systems. The same procedure can be
used for dissipative systems following the standard procedure of Bohmian mechanics. 
The resulting scaled trajectories are obtained in terms of a continuous parameter within the interval
$[0,1]$ allowing us a smooth transition from Bohmian to classical trajectories by reducing the
parameter from 1 to 0.
Arrival times and actual momentum distribution functions are also analyzed from these scaled
trajectories. The propagation of a Gaussian wave packet in a viscid medium under the presence 
of constant, linear and harmonic potentials is studied in order to simply illustrate the consequences
of this approach.  The gradual decoherence process and localization are easily visualized and
understood within this theoretical framework. 
Finally, we would like to emphasize that this approach used for conservative and dissipative systems could be considered as an alternative 
and efficient way to the WKB approximation.


\section{classical transition equation and scaled Schr\"{o}dinger equation in 
the Caldirola-Kanai model}

In studying the motion of a body in a dissipative medium, it is assumed that the friction force is a
function of its velocity. These forces are not conservative and kinetic energy is usually lost as heat.
To simplify theoretical considerations, a linear friction force in velocity is assumed.
``For slow, laminar, non-turbulent motion through a viscous fluid, the resistance is
indeed simply proportional to the speed, as can be shown at least by dimensional
arguments" \cite{Tatum-book}.
Thus, the classical equation of motion for a particle with mass $m$ in a viscid medium with friction 
$\gamma$ under the influence of a one dimensional potential $V(x)$ reads as
\begin{eqnarray} \label{eq: clas_eq_damped}
m \ddot{x} + m \gamma \dot{x} + \frac{\partial}{\partial x}V(x) &=& 0 .
\end{eqnarray}
%

%
This equation of motion   can be derived from the following 
Lagrangian \cite{Razavy-book-2005}
\begin{eqnarray*} \label{eq: class_lagrange}
\mathcal{L} &=& \left[ \frac{m}{2} \dot{x}^2 - V(x)  \right] e^{\gamma t} 
\end{eqnarray*}
where the  canonical momentum $p_c$ is given by
\begin{eqnarray} \label{eq: class_canonical momentum}
p_c &=& \frac{\partial \mathcal{L}}{\partial \dot{x}} = m \dot{x} e^{\gamma t} ,
\end{eqnarray}
which is explicitly time-dependent. Notice the difference between canonical momentum $p_c$
and kinematic one $m \dot{x} = p_c e^{-\gamma t}$.
From this Lagrangian, the so-called CK Hamiltonian is given by \cite{Razavy-book-2005}
\begin{eqnarray} \label{eq: class_Ham}
H &=& \dot{x} p_c - \mathcal{L} = \frac{ p_c ^2}{2m} e^{-\gamma t} + V(x) e^{\gamma t} .
\end{eqnarray}

The Hamiltonian operator $\hat{H}$ corresponding to the classical function (\ref{eq:
class_Ham}) can be obtained from the standard quantization rule by substituting the 
canonical momentum $p_c$ by $\frac{\hbar}{i} \frac{\partial}{\partial x}$,
\begin{eqnarray} \label{eq: KC QM-Hamiltonian}
\hat{H} &=& - \frac{\hbar^2}{2m} e^{-\gamma t} \frac{\partial^2}{\partial x^2} + 
e^{\gamma t} V(x) .
\end{eqnarray}
The commutation relation is now given by $[x, p_c] = i \hbar$ and the 
uncertainty principle $\Delta x \Delta p_c \sim \hbar $ is formally satisfied.
However, the commutation relation  of the position operator and kinematic momentum is
$i \hbar e^{-\gamma t}$, leading to the violation of the uncertainty principle. 
Notice that as long as quantities related to the momentum are not
computed, the use of the wave equation in the physical coordinate space is formally correct
\cite{SaCaRoLoMi-AP-2014}. In other words,  this violation of uncertainty principle
can be justified for friction, simulating the action of an almost macroscopic medium coupled
to the particle, since it makes the motion as time proceeds more and more predictable 
\cite{Pa-JPA-1997}.
The time-dependent Schr\"{o}dinger equation within the 
CK framework then reads  as
\begin{eqnarray} \label{eq: Sch_viscid}
i \hbar \frac{\partial}{\partial t}\psi(x, t) &=& \left[ -\frac{\hbar^2}{2m} e^{-\gamma t}
\frac{\partial^2}{\partial x^2} + V(x) e^{\gamma t}
\right] \psi(x, t) .
\end{eqnarray}
By using the polar form of the wave function as in Bohmian mechanics, 
$\psi(x, t) = R(x, t) e^{iS(x, t)/\hbar}$, one obtains
\begin{eqnarray}
\frac{\partial  R^2}{\partial t} + \frac{\partial}{\partial x}\left( R^2 \frac{1}{m} 
\frac{\partial  S}{\partial x} e^{-\gamma t} \right)  &=& 0 , \label{eq: continuity_viscid} \\
\frac{\partial  S}{\partial t} + \frac{1}{2m} \left( \frac{\partial  S}{\partial x} \right)^2 
e^{-\gamma t} + V(x) e^{\gamma t} + Q e^{-\gamma t} &=& 0 , \label{eq: HamJac_viscid}
\end{eqnarray}
where the quantum potential $ Q(x, t) $ is defined as
\begin{eqnarray} \label{eq: Qp_viscid}
Q(x, t) &=&  -\frac{\hbar^2}{2m} \frac{1}{R}\frac{\partial^2 R}{\partial x^2}   .
\end{eqnarray}
Eq. (\ref{eq: continuity_viscid}) is the continuity equation which can be written as
\begin{eqnarray*}
\frac{\partial  \rho}{\partial t} + \frac{\partial j}{\partial x} &=& 0 ,
\end{eqnarray*}
where the probability density and probability current density are 
\begin{eqnarray}
\rho(x, t) &=& R^2(x, t) = |\psi(x, t)|^2 , \label{eq: den_prob}\\
j(x, t) &=& R^2 \frac{1}{m} \frac{\partial  S}{\partial x} e^{-\gamma t}  =
\frac{\hbar}{m}~ {\text{Im}} \left( \psi^* \frac{\partial \psi}{\partial x} \right) e^{-\gamma t}
 . \label{eq: cur_den_prob}
\end{eqnarray}
Furthermore, the Bohmian velocity field is given by the guiding condition 
\begin{eqnarray} \label{eq: velocity_field}
v_B(x, t) &=& \frac{j(x, t)}{\rho(x, t)} = \frac{1}{m} \frac{\partial  S}{\partial x}  e^{-\gamma t} ,
\end{eqnarray}
from which Bohmian trajectories $x(x^{(0)}, t)$ are obtained by integrating this guidance equation,
\begin{eqnarray} \label{eq: guid_viscid}
\frac{dx}{dt} &=& v_B(x, t)\bigg|_{x=x(x^{(0)}, t)} .
\end{eqnarray}
Throughout this work, $x^{(0)}$ indicates the initial position of the particle.
It should be noted that, in the context of Bohmian mechanics, the only initial condition is the 
position of the particle since its initial momentum is specified by the phase of the wave 
function according to Eq. (\ref{eq: velocity_field}). From Eq. (\ref{eq: guid_viscid}), we have that 
\begin{eqnarray} \label{eq: acceleration_viscid}
m\ddot{x} &=& \frac{d}{dt} \left(  e^{-\gamma t} \frac{\partial  S}{\partial x}  \right) =
e^{-\gamma t} \left[ -\gamma \frac{\partial S }{\partial x} +
\left( \frac{\partial  }{\partial t} + \dot{x} \frac{\partial  }{\partial x} \right) \frac{\partial S }{\partial x}
\right] .
\end{eqnarray}
Now, using Eqs. (\ref{eq: guid_viscid}) and (\ref{eq: HamJac_viscid}),  
Eq. (\ref{eq: acceleration_viscid}) can be rewritten as
\begin{eqnarray} \label{eq: acceleration_viscid2}
m\ddot{x} + m \gamma \dot{x} + \frac{\partial}{\partial x} (V + e^{-\gamma t} Q) &=& 0 ,
\end{eqnarray}
which resembles Newton's law of motion (\ref{eq: clas_eq_damped}) except for the extra term 
$e^{-\gamma t} \partial Q / \partial x$, where $- \partial Q / \partial x$ represents the quantum force derived from the
quantum potential. Without the time exponential factor, this equation can be seen as a quantum
Langevin equation where the stochastic term given by the noise is absent.  

Following Rozen's procedure, by subtracting the non-linear term $ e^{-\gamma t} Q $ to the 
classical potential term $ e^{\gamma t} V $ in the Schr\"{o}dinger equation (\ref{eq: Sch_viscid}), 
the classical wave equation in the CK framework is expressed as
\begin{eqnarray}
i \hbar \frac{\partial}{\partial t}\psi_{\text{cl}}(x, t) &=& \left[ -\frac{\hbar^2}{2m} e^{-\gamma t}
\frac{\partial^2}{\partial x^2} + V(x) e^{\gamma t} + \frac{\hbar^2}{2m} \frac{\partial_x^2 |\psi_{\text{cl}}(x, t)|}{|\psi_{\text{cl}}(x, t)|}  e^{-\gamma t}
\right] \psi_{\text{cl}}(x, t) . \nonumber\\ \label{eq: class_wave equation}
\end{eqnarray}
%
%
We also regard classical mechanics as the limiting case of quantum mechanics. In this sense, the 
classical non-linear differential equation (\ref{eq: class_wave equation}) is the limiting case 
of the Schr\"{o}dinger linear equation (\ref{eq: Sch_viscid}). 
The last term in (\ref{eq: class_wave equation}) which is proportional to the quantum potential 
is responsible for the non-linearity. As we have shown in appendix \ref{app: class_wave}, it must 
be there in order to recover the classical Hamilton-Jacobi equation where $\hbar$ does not appear.
In this way, the classical statistical mechanics of a single particle is reformulated in a form which
 is similar to the formal structure of quantum mechanics.
Wave-particle duality is present in classical mechanics, but the wave has a purely passive role: 
the state of a classical system is still determined by its position and momentum.
In this scheme, the meaning of $\psi_{cl}$ is purely descriptive or mathematical.
On the contrary, in Bohmian mechanics, the state of a system is given by its wavefunction and 
position and the later is guided by the former. 
After Holland \cite{Holland-book-1993}, ``the quantum and the classical world are both aspects of a 
single, undivided universe. One {\it continuously} passes from one regime to the other by varying 
the effectiveness of the quantum potential." A similar approach is also used by Allori et al. 
\cite{Allori-2002}.
Throughout this work, the subindex ``cl" stands for classical motion.

A quantum-classical transition wave equation can also be introduced according 
to  \cite{RiSchMaVaBa-PRA-2014}
\begin{eqnarray}
i \hbar \frac{\partial}{\partial t}\psi_{\epsilon}(x, t) &=& \left[ -\frac{\hbar^2}{2m} e^{-\gamma t}
\frac{\partial^2}{\partial x^2} + V(x) e^{\gamma t} + (1-\epsilon) \frac{\hbar^2}
{2m} \frac{\partial_x^2 |\psi_{\epsilon}(x, t)|}{|\psi_{\epsilon}(x, t)|}  e^{-\gamma t}
\right] \psi_{\epsilon}(x, t) , \nonumber \\ \label{eq: quan_class transition}
\end{eqnarray}
where a degree of quantumness represented by $\epsilon$ (with $ 0 \leq \epsilon \leq 1 $) has 
been included. This equation provides a {\it continuous} description for the transition process 
of physical systems from purely quantum to classical regime. Thus, for  $\epsilon=0$, 
this transition equation reduces to the classical wave equation (\ref{eq: class_wave equation}) 
while, for $\epsilon=1$, it reduces to the Schr\"{o}dinger equation (\ref{eq: Sch_viscid}) in the
CK framework. 
%
%
%
%
%
%
%
%
%

%
By substituting the polar form $R_{\ep}(x, t) e^{i S_{\ep}(x, t)/ \hbar }$ of the wave function
in Eq. (\ref{eq: quan_class transition}) and after some straightforward manipulations, 
the following equations are reached 
\begin{eqnarray}
- \frac{\partial  S_{\ep}}{\partial t} \ti{\psi} &=& \frac{1}{2m} e^{-\gamma t} \left( \frac{\partial  S_{\ep}}{\partial x} \right)^2  \ti{\psi} + V(x) e^{\gamma t} \ti{\psi} -  \frac{ \ti{\hbar}^2}{2m} e^{-\gamma t} \frac{1}{R_{\ep}}
\frac{\partial^2  R_{\ep}}{\partial x^2} \ti{\psi} , \nonumber \\ \label{eq: tran_real_part2}
\\
i \ti{\hbar} \frac{\partial  R_{\ep}}{\partial t} e^{iS_{\ep}/\ti{\hbar}} &=&  -\frac{\ti{\hbar}^2}{2m } e^{-\gamma t}  \left[ \frac{2i}{\ti{\hbar}}
\frac{\partial  R_{\ep}}{\partial x} \frac{\partial  S_{\ep}}{\partial x} e^{ iS_{\ep}/\ti{\hbar} } + \frac{i}{\ti{\hbar}} \frac{\partial^2  S_{\ep}}{\partial x^2} \ti{\psi} \right] , \label{eq: tran_imag_part2}
\end{eqnarray}
where the scaled Planck's constant 
\begin{eqnarray} \label{eq: scaled Planck}
\ti{\hbar} &=& \hbar ~ \sqrt{\epsilon} ,
\end{eqnarray}
as well as the scaled wave function in polar form 
\begin{eqnarray} \label{eq: scaled_wf_polar}
\ti{\psi}(x, t) &=& R_{\ep}(x, t) e^{i S_{\ep}(x, t)/ \ti{\hbar} }
\end{eqnarray}
have been defined.
By adding Eq. (\ref{eq: tran_real_part2}) and Eq. (\ref{eq: tran_imag_part2}),  the scaled linear 
Schr\"{o}dinger equation is obtained
\begin{eqnarray} \label{eq: Scaled Sch}
i \tilde{\hbar} \frac{\partial}{\partial t}\ti{\psi}(x, t) &=& \left[ -\frac{\tilde{\hbar}^2}{2m} e^{-\gamma t}
\frac{\partial^2}{\partial x^2} + V(x) e^{\gamma t}
\right] \ti{\psi}(x, t) .
\end{eqnarray}
Thus, the nonlinear transition equation (\ref{eq: quan_class transition}) is equivalent to the 
scaled linear Schr\"{o}dinger equation (\ref{eq: Scaled Sch}) and has the same structure than 
Eq. (\ref{eq: Sch_viscid}). This will be our working equation for the scaled wave function, 
which can also be expressed in terms of the transition wave function as
\begin{eqnarray} \label{eq: scaled_transtion relation}
\ti{\psi}(x, t) &=& \psi_{\ep}(x, t) \exp \left[ \frac{i}{\hbar} \left( \frac{1}{\sqrt{\ep}} - 1  \right) S_{\ep}(x, t) \right] .
\end{eqnarray}
%

\section{Probability distribution in Bohmian  mechanics: Arrival time and actual momentum}

As is well known, the complete description of a system in Bohmian mechanics is given 
by its particle position and wave function. Quantum trajectories $x(x^{(0)}, t)$ are obtained
from the guidance equation.
Assuming that the initial distribution function for particle positions is given by the Born rule,
it is concluded by the continuity equation that such a rule holds at any time
\begin{eqnarray*}
\rho(x, t) &=& \int dx^{(0)} \rho(x^{(0)}, 0) ~ \delta \left( x - x(x^{(0)}, t) \right) .
\end{eqnarray*}
Moreover, the probability distribution function for a {\it particle property} $f$ is given by 
\cite{Le-book-2002}
\begin{eqnarray} \label{eq: dis_func}
\Pi(f) &=& \int dx^{(0)} ~\rho(x^{(0)}, 0) ~ \delta \left( f - f(x^{(0)}) \right) ,
\end{eqnarray}
$ f(x^{(0)}) $ being the value of $f$ along the quantum trajectory $ x(x^{(0)}, t) $.
Thus, for the arrival time distribution at a given location $X$, one has 
\begin{eqnarray}
\Pi_X(t) &=& \int dx^{(0)} \rho(x^{(0)}, 0) ~ \delta \left( t - T_X(x^{(0)}) \right) ,
\end{eqnarray}
with $T_X(x^{(0)})$ giving the time at which particles arrive at the detector location $X$
and the mean arrival time is given by
\begin{eqnarray}
\tau(X) &=& \int_0^{\infty} dt ~ t ~\Pi_X(t)  \label{eq: tau_def1} \\
&=& \int_0^{\infty} dt ~ t \int dx^{(0)} \rho(x^{(0)}, 0) ~ \delta \left( t - T_X(x^{(0)}) \right)  \nonumber \\
&=& \int dx^{(0)} \rho(x^{(0)}, 0) \int_0^{\infty} dt ~ t ~\delta \left( t - T_X(x^{(0)}) \right) \nonumber \\
&=& \int dx^{(0)} \rho(x^{(0)}, 0) ~ T_X(x^{(0)}) .  \label{eq: tau_def2}
\end{eqnarray}
By using the properties of the Dirac delta function and non-crossing property of Bohmian trajectories,
Leavens \cite{Le-book-2002} obtained
\begin{eqnarray} \label{eq: ar_dis_pcd}
\Pi_X(t) &=& \frac{|j(X, t)|}{ \int_0^{\infty} dt'|j(X, t')| }
\end{eqnarray}
for the {\it intrinsic} (apparatus independent) arrival time distribution for those particles that 
actually arrive at the detector location $X$. It  should be notice that this relation is naturally obtained 
within the Bohmian mechanics. There are several theoretical treatments for the arrival 
time distribution within the standard theory of quantum mechanics
due to the absence of a clear treatment to include time observables into its formalism
(for a review see, for example, \cite{MuLe-PR-2000}).
As pointed out in this review, different theoretical methods approach to Eq. (\ref{eq: ar_dis_pcd}) 
at asymptotic distances and times from the particle source and existing scatterers.

On the other hand, the Bohmian momentum filed is given by $ p(x, t) = m ~ \dot{x}(x, t) $. 
Thus, using Eq. (\ref{eq: dis_func}) one has
\begin{eqnarray} \label{eq: mom_dis}
\Pi(p, t) &=& \int dx^{(0)} ~\rho(x^{(0)}, 0) ~ \delta \left( p -  m ~ \dot{x}(x, t) \bigg|_{ x = x(x^{(0)}, t) } \right)
\end{eqnarray}
for the {\it actual} momentum \cite{Holland-book-1993} distribution function.


\section{Propagation of a Gaussian wave packet. Scaled trajectories}

We will consider evolution of the scaled Gaussian wave packet
\begin{eqnarray} \label{eq: inital_Gauss}
\tilde{\psi}(x, 0) &=& \frac{1}{ (2\pi \sigma_0^2)^{1/4} } \exp \left[ -\frac{(x-x_0)^2}{4\sigma_0^2} + \frac{i}{\tilde{\hbar}}
p_0(x-x_0) \right]
\end{eqnarray}
in a viscid medium for simple potential functions  using the corresponding 
propagator for the scaled time dependent Schr\"odinger equation  (\ref{eq: Scaled Sch}) 
given in \ref{app: propag}. Here,
$\sigma_0$, $x_0$ and $p_0$ are the initial width, position and momentum, respectively.
In the free motion case,  the time dependent scaled wave function reads as
\begin{eqnarray} \label{eq: scaled psi_G(t)}
\tilde{\psi}(x, t) &=& \frac{1}{ (2\pi  \tilde{s}_t^2)^{1/4} } \exp \left[ -\frac{(x-x_t)^2}{4\sigma_0  \tilde{s}_t} + \frac{i}{\tilde{\hbar}} ~p_t ~e^{\gamma t} (x-x_t) + \frac{i}{\tilde{\hbar}} \mathcal{A}_{\text{cl},t} \right]  ,
\end{eqnarray}
$ \tilde{s}_t $ being  a complex function of time and friction
$\gamma$; its modulus is the width of the wave packet. Quantities $x_t$, $p_t$ and 
$ \mathcal{A}_{\text{cl},t} $ denote respectively position of the center, associated momentum 
and classical action along the trajectory $x_t$ with
\begin{eqnarray} \label{eq: action_class}
\mathcal{A}_{\text{cl},t} &=& \int_0^t dt' \mathcal{L}_{t'} = \int_0^t dt' \left( \frac{1}{2} m \dot{x}_{t'}^2 + V(x_{t'}) \right) e^{\gamma t'} .
\end{eqnarray}
The center of the wave packet follows a classical trajectory $x_t$  and coincides with the 
expectation value $ \langle x \rangle(t) $  according to the Ehrenfest theorem.

The probability and probability current densities are respectively given by
\begin{eqnarray}
\ti{\rho}(x, t) &=& \frac{1}{ \sqrt{2\pi \ti{\sigma}_t^2 }} \exp \left[ - \frac{(x-x_t)^2}{2\ti{\sigma}_t^2} \right] ,
\label{eq: rho_general} \\
\ti{j}(x, t) &=& \frac{\ti{\hbar}}{m}~ {\text{Im}} \left( \ti{\psi}^* \frac{\partial \ti{\psi}}{\partial x} \right) e^{-\gamma t} =
\frac{\ti{\hbar}}{m}
\left[ \frac{x-x_t}{2\sigma_0 \ti{\sigma}_t^2} ~ {\text{Im}}( \ti{s}_t ) ~ e^{-\gamma t} + \frac{p_t}{ \ti{\hbar} }  \right]  \ti{\rho}(x, t) ,
\nonumber \\
\label{eq: j_general}
\end{eqnarray}
with
\begin{eqnarray} \label{eq: sigmaX_general}
\tilde{\sigma}_t &=& | \ti{s}_t |     
\end{eqnarray}
and the coefficient of $\ti{\rho}(x, t)$ in Eq. (\ref{eq: j_general}) is the scaled velocity field
\begin{eqnarray} \label{eq: BM_velocity field} 
v_s(x, t) &=& \frac{\ti{\hbar}}{2m \sigma_0} \frac{ e^{-\gamma t}{\text{Im}}( \ti{s}_t  ) }{ \ti{\sigma}_t^2 } ( x-x_t ) + \frac{p_t}{ m }   .
\end{eqnarray}
Scaled trajectories are then expressed  from the guidance equation (\ref{eq: guid_viscid}) as
\begin{eqnarray*}
\frac{d}{dt}(x-x_t) &=& \frac{\ti{\hbar}}{2m \sigma_0} \frac{ e^{-\gamma t} {\text{Im}}(  \ti{s}_t ) }{ \ti{\sigma}_t^2 }
( x-x_t ) ,
\end{eqnarray*}
which has the solution,
\begin{eqnarray*}
\ln \left( \frac{x(x^{(0)}, t) - x_t}{x^{(0)} - x_0 } \right) &=& \frac{\ti{\hbar}}{2m \sigma_0}
\int_0^t dt' \frac{ e^{-\gamma t'} {\text{Im}}( \ti{s}_{t'} ) }{ \ti{\sigma}_{t'}^2 } .
\end{eqnarray*}
or written differently 
\begin{eqnarray} \label{Bohm_traj_generalform}
x(x^{(0)}, t) &=& x_t + g(t) ~ \frac{\ti{\sigma}_t}{\sigma_0} ~ (x^{(0)} - x_0) ,
\end{eqnarray}
with
\begin{eqnarray} \label{eq: au_g_function}
g(t) &=& \frac{\sigma_0}{\ti{\sigma}_t}~\exp \left[ \frac{\ti{\hbar}}{2m \sigma_0} \int_0^t dt' \frac{ e^{-\gamma t'} {\text{Im}}(\ti{s}_{t'}) }{ \ti{\sigma}_{t'}^2 } \right] .
\end{eqnarray}
Eq. (\ref{Bohm_traj_generalform}) shows that if the particle is initially located at the 
center of the wave packet, it remains there forever, following its classical path $x_t$.
The structure of this equation presents a general {\it dressing} scheme \cite{salva3}
where the scaled trajectory consists of a classical trajectory (a particle property) plus 
a term involving the width of the wave packet (a wave property). Bohmian and classical trajectories 
are obtained for the two extreme values of the continuous parameter $\epsilon=1,0$, respectively

On the other hand, from Eq. (\ref{Bohm_traj_generalform}) one also obtains
\begin{eqnarray} \label{eq: Bohm_mom_generalform}
m ~ \dot{x}(x^{(0)}, t) &=& p_t + m~f(t)~\frac{ \dot{ \tilde{\sigma}}_t }{\sigma_0}~(x^{(0)} - x_0)
\end{eqnarray}
for the momentum along the quantum trajectory, where we have introduced the
auxiliary function
\begin{eqnarray} \label{eq: au_f_function}
f(t) &=& g(t) + \frac{ \tilde{\sigma}_t }{ \dot{ \tilde{\sigma}}_t  } \dot{g}(t)
\end{eqnarray}
for simplicity. An alternative way to obtain the momentum along a trajectory is substituting 
Eq. (\ref{Bohm_traj_generalform}) into Eq. (\ref{eq: BM_velocity field}).
Thus, from Eq. (\ref{eq: mom_dis}), the actual momentum distribution can be
expressed as
\begin{eqnarray}
\ti{\Pi}(p, t) &=& \frac{1}{ \sqrt{2\pi \sigma_0^2} } \int dx^{(0)} ~ \exp \left[ -\frac{(x^{(0)}-x_0)^2}{2\sigma_0^2} \right]
~ \delta \left( p - p_t - mf(t) \frac{\dot{ \tilde{\sigma}}_t}{\sigma_0} (x^{(0)}-x_0)  \right)
\nonumber \\
&=&
\frac{1}{ \sqrt{2\pi \sigma_0^2} } \int dy ~ \exp \left[ -\frac{y^2}{2\sigma_0^2} \right]
~ \delta \left( mf(t) \frac{\dot{ \tilde{\sigma}}_t}{\sigma_0} y + p_t - p )  \right)
\nonumber
\\
&=&
\frac{1}{ \sqrt{2\pi \tilde{\Sigma}_t^2} } ~ \exp \left[ -\frac{(p-p_t)^2}{2\tilde{\Sigma}_t^2} \right] ,
\label{eq: mom_dis__generalform}
\end{eqnarray}
with
\begin{eqnarray} \label{eq: sigmaP_general}
\tilde{\Sigma}_t &=& mf(t) \dot{\tilde{\sigma}}_t
\end{eqnarray}
being the width of the momentum distribution. 
In \ref{ap: four_trans}, the Fourier transform of the free damped Gaussian wave packet 
whose module-squared is the quantum distribution function for the momentum is derived.
For the product of widths, we have
\begin{eqnarray} \label{eq: widths_product_general}
\tilde{\sigma}_t \tilde{\Sigma}_t &=& \frac{m}{2} f(t) \frac{d}{dt} ~ \tilde{\sigma}_t^2 ~.
\end{eqnarray}
Now, let us consider some special cases.

çç

\subsection{Free scaled Gaussian wave packet in non-viscid medium}

For a free particle, $V(x) = 0$, in a non-viscid medium, $\gamma=0$, the results are similar to 
those found in \cite{Holland-book-1993} for a nonscaled wave function. One has
\begin{eqnarray} \label{eqs: free_prop}
\begin{cases}
\tilde{s}_t = \sigma_0 \left( 1+i\frac{ \sqrt{\epsilon} \hbar t}{2m\sigma_0^2} \right) , \\
x_t = x_0 + \frac{p_0}{m} t ,   \\
p_t = m \dot{x}_t = p_0 ,  \\
\mathcal{A}_{\text{cl},t} = \frac{p_0^2}{2m} t ,
\end{cases}
\end{eqnarray}
the auxiliary functions being given by
\begin{eqnarray} \label{eqs: free_au_gf_functions}
\begin{cases}
g(t) = 1  ,   \\
f(t) = 1 .
\end{cases}
\end{eqnarray}
For the widths in configuration and actual momentum spaces, we have
\begin{eqnarray} \label{eqs: free_widths}
\begin{cases}
\tilde{\sigma}_t = \sigma_0 \sqrt{ 1 + \frac{ \epsilon \hbar^2 t^2}{4 m^2 \sigma_0^4} }  ,   \\
\tilde{\Sigma}_t = \frac{ \epsilon \hbar^2 t}{4 m \sigma_0^2 \tilde{\sigma}_t} 
\end{cases}
\end{eqnarray}
and at asymptotic times, $ t \rightarrow \infty $, 
\begin{eqnarray} \label{eqs: widths_largetime_free}
\begin{cases}
\tilde{\sigma}_{\infty}    \approx \frac{ \sqrt{\epsilon} ~ \hbar }{2 m \sigma_0 } ~ t  , \\
\tilde{\Sigma}_{\infty} \approx \frac{ \sqrt{\epsilon} ~ \hbar }{2 \sigma_0} \equiv \sigma_p .
\end{cases}
\end{eqnarray}
Thus, in this limit, one easily recognizes that the actual momentum distribution coincides 
with the usual quantal one
\begin{eqnarray} \label{eq: quantal_p_dis}
\lim_{t \rightarrow \infty} \tilde{\Pi}(p, t) &=& \frac{1}{ \sqrt{2\pi \ti{\sigma}_p^2 }}
\exp \left[ - \frac{ ( p-p_0)^2}{2\ti{\sigma}_p^2} \right]
\equiv
| \ti{\phi}(p, 0)|^2 ,
\end{eqnarray}
where $\ti{\phi}(p, 0)$ is the Fourier transform of the initial wave function 
(\ref{eq: inital_Gauss}) which is evaluated in \ref{ap: four_trans}.

Scaled trajectories follow the dressing scheme 
\begin{eqnarray} \label{eq: BM traj_free}
x(x^{(0)}, t) &=& x_0 + \frac{p_0}{m} t + (x^{(0)} - x_0) 
\sqrt{ 1 + \frac{ \epsilon \hbar^2 t^2}{4 m^2 \sigma_0^4} } ,
\end{eqnarray}
using  Eqs. (\ref{eqs: free_prop}), (\ref{eqs: free_au_gf_functions}) and (\ref{eqs: free_widths}) 
in  Eq. (\ref{Bohm_traj_generalform}). 
Eq. (\ref{eq: BM traj_free}) has also been obtained in \cite{Ch-AP-2016}.

The $\epsilon$ parameter is accompanying the wave feature as a factor. Bohmian trajectories 
are obtained from these scaled trajectories 
for $\epsilon = 1$. 

In the classical regime, when $\epsilon = 0$, from Eqs. (\ref{eq: rho_general}), 
(\ref{eq: j_general}) and (\ref{eq: mom_dis__generalform}), one has
\begin{eqnarray}
\rho_{\text{cl}}(x, t) &=& \frac{1}{ \sqrt{2\pi \sigma_0^2} } \exp \left[ -\frac{(x - x_0 - \frac{p_0}{m} t )^2}{2\sigma_0^2} \right] ,
\label{eq: class_x_dis_free} \\
j_{\text{cl}}(x, t) &=& \frac{p_0}{m} \rho_{\text{cl}}(x, t) , \label{eq: class_j_free} \\
\Pi_{\text{cl}}(p, t) &=& \delta(p-p_0) , \label{eq: class_p_dis_free}
\end{eqnarray}
which shows that we have a Gaussian wave packet in configuration space which moves with 
constant velocity $p_0/m$ and maintains its width. However, the actual momentum has a Dirac 
delta distribution around $p = p_0$, i.e., all particles in the ensemble have the same momentum 
$p_0$, which is apparent from Eq. (\ref{eq: Bohm_mom_generalform}) in the classical limit where 
$\dot{\ti{\sigma}}_t = 0$; and noting that $p_t=p_0$.
The arrival time distribution at the detector position $X$ has a simple form in this classical limit,
\begin{eqnarray} \label{eq: class_ar_dis}
\Pi_{\text{cl}, X}(t) &=& \sqrt{\frac{ 2 p_0^2 }{\pi m^2 \sigma_0^2} } \exp \left[ -\frac{p_0^2}{2m^2\sigma_0^2}
\left( t - \frac{m}{p_0}(X-x_0)  \right)^2 
\right]   ,
\end{eqnarray}
showing a Gaussian shape with a center at $m(X-x_0)/p_0$.
From  Eq. (\ref{Bohm_traj_generalform}), the classical arrival time at the detector 
position $X$ is given by
\begin{eqnarray} \label{eq: class_ar_t}
T_{X, \text{cl}}(x^{(0)}) &=& \frac{m}{p_0}(X-x^{(0)}) ,
\end{eqnarray}
and the classical mean arrival time is finally expressed as 
\begin{eqnarray} \label{eq: class_tau}
\tau_{\text{cl}}(X) &=& \frac{m}{p_0}(X - \langle x \rangle_0  ) = \frac{m}{p_0}(X - \sigma_0  )
\end{eqnarray}
from Eqs. (\ref{eq: class_x_dis_free}) and (\ref{eq: class_ar_t}).

\subsection{Free scaled Gaussian wave packet in a viscous medium}

In this case, the free evolution of the Gaussian wave packet (\ref{eq: inital_Gauss}) in a 
viscid medium is considered.  From the free propagator of the Schr\"{o}dinger equation which 
we have derived  in \ref{app: propag} (Eqs.(\ref{eq: wavet_propag_wave0}), 
(\ref{eq: free_CK_propagator})  and (\ref{eq: inital_Gauss})), we have
\begin{eqnarray}
\ti{\psi}(x, t) &=& \sqrt{ \frac{m\gamma}{2\pi i \ti{\hbar} (1-e^{-\gamma t})} }
\frac{1}{ (2\pi \sigma_0^2)^{1/4} }
\int dx' \exp \left[ \frac{i m}{2 \ti{\hbar} } \frac{\gamma}{1-e^{-\gamma t}} (x-x')^2 -\frac{(x'-x_0)^2}{4\sigma_0^2} + \frac{i}{\tilde{\hbar}} p_0(x'-x_0) \right]
\nonumber \\
&=&
\sqrt{ \frac{m\gamma}{2\pi i \ti{\hbar} (1-e^{-\gamma t})} } \frac{1}{ (2\pi \sigma_0^2)^{1/4} } 
\sqrt{ \frac{\pi}{d_1} } e^{ d_2^2 / 4 d_1 + d_3} , \label{eq: wavet_propag_wave0}
\end{eqnarray}
where
\begin{eqnarray*}
d_1 &=& \frac{1}{ 4 \sigma_0^2 } - \frac{i m}{2 \ti{\hbar} } \frac{\gamma}{1-e^{-\gamma t}} , \\
d_2 &=& - \frac{i m}{ \ti{\hbar} } \frac{\gamma}{1-e^{-\gamma t}} x + \frac{1}{ 2 \sigma_0^2 } x_0 + \frac{i}{\tilde{\hbar}} p_0 , \\
d_3 &=& \frac{i m}{2 \ti{\hbar} } \frac{\gamma}{1-e^{-\gamma t}} x^2 - \frac{1}{ 4 \sigma_0^2 } x_0^2 - \frac{i}{\tilde{\hbar}} p_0 x_0 .
\end{eqnarray*}
After straightforward but lengthy calculations, one arrives at the shape (\ref{eq: scaled psi_G(t)}) 
for the wave function but now with
\begin{eqnarray} \label{eqs: free_prop_viscid}
\begin{cases}
\ti{s}_t = \sigma_0 \left( 1 + i \frac{ \sqrt{\epsilon} \hbar}{2m\sigma_0^2} \frac{1-e^{-\gamma t}}{\gamma} \right) ,  \\
x_t = x_0 + \frac{p_0}{m} \frac{1-e^{-\gamma t}}{\gamma} ,  \\
p_t = m \dot{x}_t = p_0 e^{-\gamma t} ,  \\
\mathcal{A}_{\text{cl},t} = \frac{p_0^2}{2m} \frac{1-e^{-\gamma t}}{\gamma} ,
\end{cases}
\end{eqnarray}
instead of Eqs. (\ref{eqs: free_prop}).
From the expression $ e^{\gamma t} p_t = e^{\gamma t} m \dot{x}_t $ for the canonical 
momentum, it has been argued that the CK Hamiltonian describes a particle of exponentially
growing mass, $m(t) = m e^{\gamma t}$ \cite{Gr-JMP-1979}.
For the auxiliary function, we have that 
\begin{eqnarray*}
g(t) &=& \frac{\sigma_0}{\tilde{\sigma}_t} \exp \left[ \frac{1}{2} \log \left( \frac{\tilde{\sigma}_t^2}{\sigma_0^2} \right) \right] = 1 ,
\end{eqnarray*}
by noting  that
\begin{eqnarray*}
\frac{d}{dt} \tilde{\sigma}_t^2 &=& \frac{\tilde{\hbar}}{m \sigma_0} e^{-\gamma t} \text{Im}(\tilde{s}_t)
\end{eqnarray*}
and  $f(t)=1$ again, meaning that Eqs. (\ref{eqs: free_au_gf_functions}) are still valid.
Analogously, for the widths of the distributions in configuration and actual momentum, one has
\begin{eqnarray} \label{eqs: free_widths_viscid}
\begin{cases}
\tilde{\sigma}_t = \sigma_0 \sqrt{ 1 + \frac{ \epsilon \hbar^2 }{4 m^2 \sigma_0^4} \left( \frac{1-e^{-\gamma t}}{\gamma} \right)^2 } ,   \\
\tilde{\Sigma}_t = \frac{ \epsilon \hbar^2}{4 m \sigma_0^2 \tilde{\sigma}_t} \frac{ e^{-\gamma t}(1-e^{-\gamma t}) }{\gamma} .
\end{cases}
\end{eqnarray}

Scaled trajectories within the dressing scheme are now  given by 
\begin{equation} \label{eq: BM traj_free_viscid}
x(x^{(0)}, t) = x_0 + \frac{p_0}{m} \frac{1-e^{-\gamma t}}{\gamma} + (x^{(0)} - x_0)
\sqrt{ 1 + \frac{ \epsilon \hbar^2 }{4 m^2 \sigma_0^4} \left( \frac{1-e^{-\gamma t}}{\gamma} \right)^2 } ,
\end{equation}
using  Eqs. (\ref{eqs: free_prop_viscid}), (\ref{eqs: free_au_gf_functions}) and 
(\ref{eqs: free_widths_viscid}) in  Eq. (\ref{Bohm_traj_generalform}). For
$\epsilon > 0$, these trajectories correspond to an intermediate regime,  leading to the
Bohmian trajectories when $\epsilon = 1$. Again, this parameter appears as a factor in the 
non-classical part. This intermediate regime is critical to better understand the gradual decoherence
process in this global dynamics.
For a given value of friction $\gamma$ and at a given instant of time, widths behave like
\begin{eqnarray*}
\tilde{\sigma}_t &\sim & \sqrt{1 + c~ \epsilon } , \\
\tilde{\Sigma}_t &\sim & \frac{\epsilon}{\sqrt{1 + c~ \epsilon }} ,
\end{eqnarray*}
where $c$ is a constant. Thus, in the transition from the quantum to classical regime, 
$\epsilon \rightarrow 0$, both widths decrease from its maximum value in the quantum regime 
to the minimum one in the classical regime.

In viscid media, $ \gamma \neq 0 $, at $t=0$, $\tilde{\Sigma}_0 = 0$ which shows that all 
particles of the ensemble initially have the same velocity $p_0/m$.
As $ t \rightarrow \infty $, from Eqs. (\ref{eqs: free_widths_viscid}), (\ref{eq: rho_general}) 
and (\ref{eq: mom_dis__generalform}) one has
\begin{eqnarray} \label{eqs: large time limit eqs}
\tilde{\sigma}_{\infty}   & \approx & \sigma_0 \sqrt{ 1 + \frac{ \epsilon \hbar^2}{4 m^2 \sigma_0^4 \gamma^2} } , \\
\tilde{\Sigma}_{\infty} & \approx & 0 , \\
\ti{\rho}(x, \infty)  & \approx & \frac{1}{ \sqrt{2\pi \ti{\sigma}_{\infty}^2 }}
\exp \left[ - \frac{ \left( x-x_0- \frac{p_0}{m \gamma} \right)^2}{2\ti{\sigma}_{\infty}^2} \right] ,\\
\tilde{\Pi}(p, t) & \rightarrow & \delta(p) ,
\end{eqnarray}
showing that at the end, the distribution function in the position representation stops at the
point $x_0 + p_0/m \gamma$, and the distribution function for the actual momentum takes the form 
of the Dirac delta around zero momentum, revealing all particles of the ensemble stop at the end.
In other words, the wave packet becomes localized, that is, motionless and with the spreading
being frozen. 
For comparison, we provide the momentum space representation of the scaled wave function and 
the corresponding distribution function in \ref{ap: four_trans}.
Moreover, from Eq. (\ref{eq: BM traj_free_viscid}), one sees that particles stop (localization) 
finally at
\begin{eqnarray} \label{eq: Bohm_stop_point}
x(x^{(0)}, \infty) &=& x^{(0)} + \frac{p_0}{m\gamma} + (  x^{(0)}-x_0 ) \sqrt{ 1 + \frac{\epsilon \hbar^2}{4 m^2 \sigma_0^4 \gamma^2} } .
\end{eqnarray}
and to arrive at the detector location $X$, the detector must be placed at $ X \leq x(x^{(0)}, \infty) $.

In the classical limit, $ \epsilon \rightarrow 0 $, one has
\begin{eqnarray} \label{eq: widths_cl}
\begin{cases}
\ti{\sigma}_t  \rightarrow  \sigma_0 ,\\
\tilde{\Sigma}_t  \rightarrow  0 ,
\end{cases}
\end{eqnarray}
and from Eqs. (\ref{eq: rho_general}) and (\ref{eq: mom_dis__generalform}) the classical 
distribution functions read as
\begin{eqnarray}
\rho_{\text{cl}}(x, t) &=& \frac{1}{ \sqrt{2\pi \sigma_0^2 }}
\exp \left[ - \frac{ \left( x - x_0 - \frac{p_0}{m} \frac{1-e^{-\gamma t}}{\gamma} \right)^2}{2\sigma_0^2} \right] ,
\\
\Pi_{\text{cl}}(p, t) &=& \delta \left( p - p_0 e^{-\gamma t} \right) .
\end{eqnarray}
These relations show that the classical distribution function in configuration space retains its 
Gaussian shape during the  motion with velocity $p_0 e^{-\gamma t} / m$, while the momentum
distribution is again a delta function, with all particles of the classical ensemble having the 
same momentum $ p_0 e^{-\gamma t} $. Furthermore,
from  Eq. (\ref{eq: BM traj_free_viscid}) one obtains
\begin{eqnarray} \label{eq: class_traj_viscid}
x_{\text{cl}}(x^{(0)}, t) &=& x^{(0)} + \frac{p_0}{m\gamma} \left( 1- e^{-\gamma t}\right)
\end{eqnarray}
for the classical trajectory and 
\begin{eqnarray} \label{eq: class_ar_t_viscid}
T_{X, \text{cl}}(x^{(0)}) &=& -\frac{1}{\gamma} \ln \left[ 1 - \frac{m\gamma}{p_0} (X-x^{(0)})   \right]
\end{eqnarray}
for the arrival time at the detector location $X$ which by the same reasoning as above, 
one should not worry about the argument of the $\ln$-function and 
positivity of $T_{X, \text{cl}}(x^{(0)})$.

All of results of this subsection reduce to the corresponding ones in the previous subsection in the 
non-viscid limit. The corresponding scaled trajectories are identical to those of 
Ref. \cite{SaCaRoLoMi-AP-2014} obtained from a given anzat for the wave function when 
$\epsilon =0,1$.

\subsection{Propagation of a scaled Gaussian wave packet in a damped linear potential}

Let us consider now the propagation of the wave packet (\ref{eq: inital_Gauss}) in viscid media 
but in the presence of an external linear potential
\begin{eqnarray} \label{eq: linear_pot}
V(x) &=& m a ~ x ,
\end{eqnarray}
$a$ being a constant acceleration.
The corresponding propagator of the scaled Schr\"{o}dinger equation (\ref{eq: Scaled Sch}) 
is given by Eq. (\ref{eq: linear_CK_propagator}) with $\tilde{\hbar}$ instead of $\hbar$. So, by 
using Eq. (\ref{eq: wavet_propag_wave0}), the time evolution of the wave function is obtained 
by straightforward but lengthy calculations as Eq. (\ref{eq: scaled psi_G(t)}). The resulting equations 
are
\begin{eqnarray} \label{eqs: linear_prop_viscid}
\begin{cases}
\ti{s}_t = \sigma_0 \left( 1 + i \frac{ \sqrt{\epsilon} \hbar }{2m\sigma_0^2} \frac{1-e^{-\gamma t}}{\gamma} \right) ,  \\
x_t = x_0 + \frac{p_0}{m} \frac{1-e^{-\gamma t}}{\gamma} - a \frac{ \gamma t - 1 + e^{-\gamma t} }{ \gamma^2 } ,  \\
p_t = m \dot{x}_t = p_0 e^{-\gamma t} - m a \frac{1-e^{-\gamma t}}{\gamma} ,  \\
\mathcal{A}_{\text{cl},t} = \frac{p_0^2}{2m} \frac{1-e^{-\gamma t}}{\gamma} - a \left[  p_0 \frac{-2+e^{-\gamma t} + e^{\gamma t}}{\gamma^2}  + m x_0 \frac{ -1+e^{\gamma t} }{\gamma} \right] + a^2 m \frac{ 4 + ( 2 \gamma t -3 ) e^{\gamma t} - e^{-\gamma t} }{\gamma^3} .
\end{cases}
\end{eqnarray}
A direct comparison with Eqs. (\ref{eqs: free_prop_viscid}) shows that $\ti{s}_t$ has the same 
expression for both propagation in zero and in constant force field. Thus, Eqs. 
(\ref{eqs: free_au_gf_functions}) are still valid and widths are given by Eq. 
(\ref{eqs: free_widths_viscid}).
On the other hand, scaled trajectories within the dressing scheme are given by 
\begin{eqnarray}
x(x^{(0)}, t) &=& x_0 + \frac{p_0}{m} \frac{1 - e^{-\gamma t}}{\gamma} - a \frac{\gamma t - 1 + e^{-\gamma t} }{\gamma^2}
\nonumber \\
&+& (x^{(0)} - x_0)
\sqrt{ 1 + \frac{ \epsilon \hbar^2 }{4 m^2 \sigma_0^4} \left( \frac{1-e^{-\gamma t}}{\gamma} \right)^2 } , \label{Bohm_traj_linear}
\end{eqnarray}
when using Eqs. (\ref{eqs: linear_prop_viscid}), (\ref{eqs: free_au_gf_functions}) and 
(\ref{eqs: free_widths_viscid}) in Eq.(\ref{Bohm_traj_generalform}). This equation shows 
that particles, just as the classical ones, move with constant velocity $- a / \gamma$ 
at large times, $t \rightarrow \infty$. Furthermore, the probability current density at the detector 
location  $X$, non-normalized arrival time distribution, is given by
\begin{eqnarray}
\ti{j}(X, t) &=&
\left[ \frac{\epsilon \hbar^2}{4 m^2 \sigma_0^2 \ti{\sigma}_t^2} \frac{e^{-\gamma t}(1- e^{-\gamma t})}{\gamma} (X-x_t) + \frac{p_t}{m} \right]
\nonumber \\
& \times &
 \frac{1}{ \sqrt{2\pi \ti{\sigma}_t^2 }}
\exp \left[ - \frac{(X-x_t)^2}{2\ti{\sigma}_t^2} \right] ,
\label{eq: PCD_linear}
\end{eqnarray}
from Eqs. (\ref{eq: rho_general}), (\ref{eq: j_general}) and (\ref{eqs: free_widths_viscid}).
 In the classical limit established by Eq. (\ref{eq: widths_cl}) we 
obtain that
\begin{eqnarray} \label{eq: PCD_linear_cl}
j_{\text{cl}}(X, t) &=&
 \frac{p_t}{m} \frac{1}{ \sqrt{2\pi \sigma_0^2 }}
\exp \left[ - \frac{(X-x_t)^2}{2 \sigma_0^2} \right] ,
\end{eqnarray}
which does {\it not} have a Gaussian shape in time.

\begin{figure}
\centering
\includegraphics[width=10cm,angle=0]{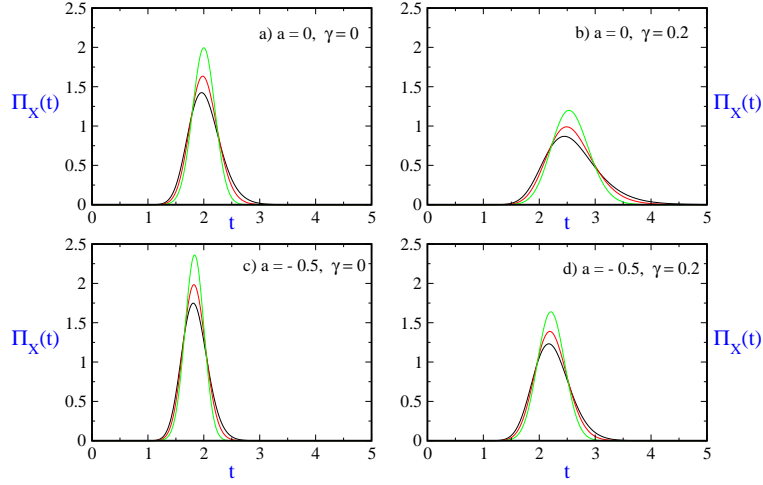}
\caption{(Color online)
Arrival time distributions for two different values of $\gamma$ and $a$ and three regmes: 
$\epsilon = 1$ (Bohmian, black curve), $\epsilon = 0.5$ (intermediate, red curve), 
$\epsilon = 0$ (classical, green curve).}
\label{fig: arrival_dis}
\end{figure}

\begin{figure}[h]
\centering
\includegraphics[width=10cm,angle=-90]{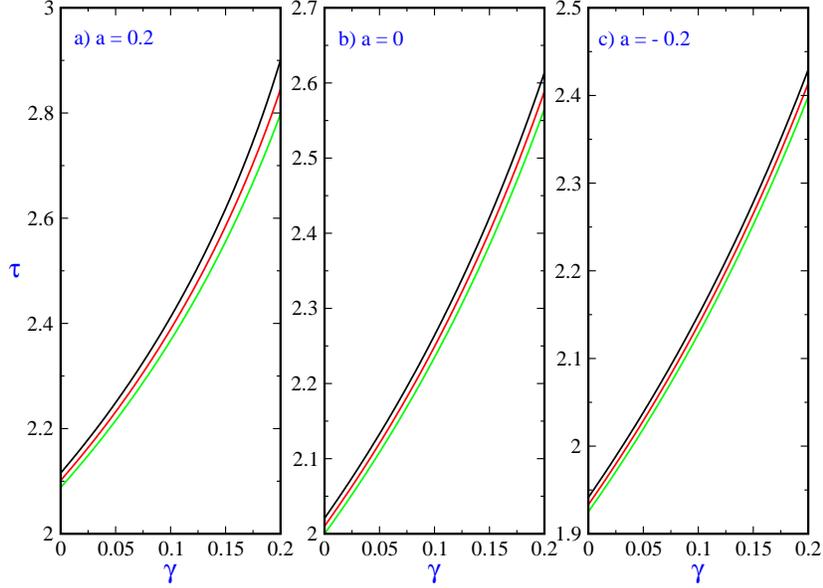}
\caption{(Color online)
Mean arrival times versus $\gamma$ for different values of $\epsilon$ (regimes) and  $a$: 
$\epsilon = 1$ (Bohmian, black curve), $\epsilon = 0.5$ (intermediate, red curve), 
$\epsilon = 0$ (classical, green curve).}
\label{fig: mean_arrival}
\end{figure}

It is very instructive to analyze in more detail this dissipative dynamics within the scaled 
wave function framework or transition quantum-classical regime.
In this regard, we work in a system of units where $\hbar =1$ and $m=1$.  We place the detector 
at $ X = 0 $. Parameters of the initial Gaussian wave packet are chosen to be $ \sigma_0 = 1 $; 
and $ x_0 = - 10 $ and $ p_0 = 5 $. Seven scaled trajectories are calculated for each 
dynamical regime considered: $\epsilon = 1$, Bohmian;  $\epsilon = 0.5$, intermediate; 
$\epsilon = 0$, classical.
As pointed above, arrival times of those particles that {\it actually} reach the detector are given 
by the modulus of the probability current density.
For computations, one has to make sure that the denominator of Eq. (\ref{eq: ar_dis_pcd}) is 
non-zero implying that the number of particles in the ensemble reaching the detector is not zero.
Due to the equivariance principle, particles distribute according to the Born rule. 
The non-crossing property of Bohmian paths implies that if a particle in the far left tail of the 
Gaussian packet (for example, $x^{(0)} = x_0 - 5\sigma_0$) reaches the detector, all other 
particles have certainly reached before.
If we want that all particles arrive at the detector, one must choose the friction coefficient such 
that $\gamma \le 0.22$ in the quantum regime, $\epsilon=1$, and $\gamma \le 0.33$ for the 
classical one, $\epsilon=0$.
In Figure \ref{fig: arrival_dis}, arrival time distributions for the free 
particle and a uniform accelerating ($a<0$ in Eq. (\ref{eq: linear_pot})) 
force field are showed. Two different values of $\gamma$ and $a$ are used for each regime: 
$\epsilon = 1$ (Bohmian, black curve), $\epsilon = 0.5$ (intermediate, red curve), 
$\epsilon = 0$ (classical, green curve). From this figure, one observes that the maximum of the 
distribution is shifted towards longer times when passing from the quantum to classical regime, 
while its width (full width at half maximum) becomes narrower.
The maximum locates at shorter times and the distribution becomes narrower for an 
{\it accelerating} force compared to the free case, revealing that particles arrive sooner at the 
detector location. The difference between arrival times of different particles of the ensemble 
becomes smaller.
Figure \ref{fig: mean_arrival} shows that mean arrival time increases with friction
$\gamma$ for three different accelerations (the same color code is used for the three dynamical
regimes than in Figure \ref{fig: arrival_dis}). This time is always shorter for the classical regime. 
This behavior is a clear manifestation of the non-Gaussian shape before alluded.
%
%
%

\begin{figure}[h]
\centering
\includegraphics[width=10cm,angle=-90]{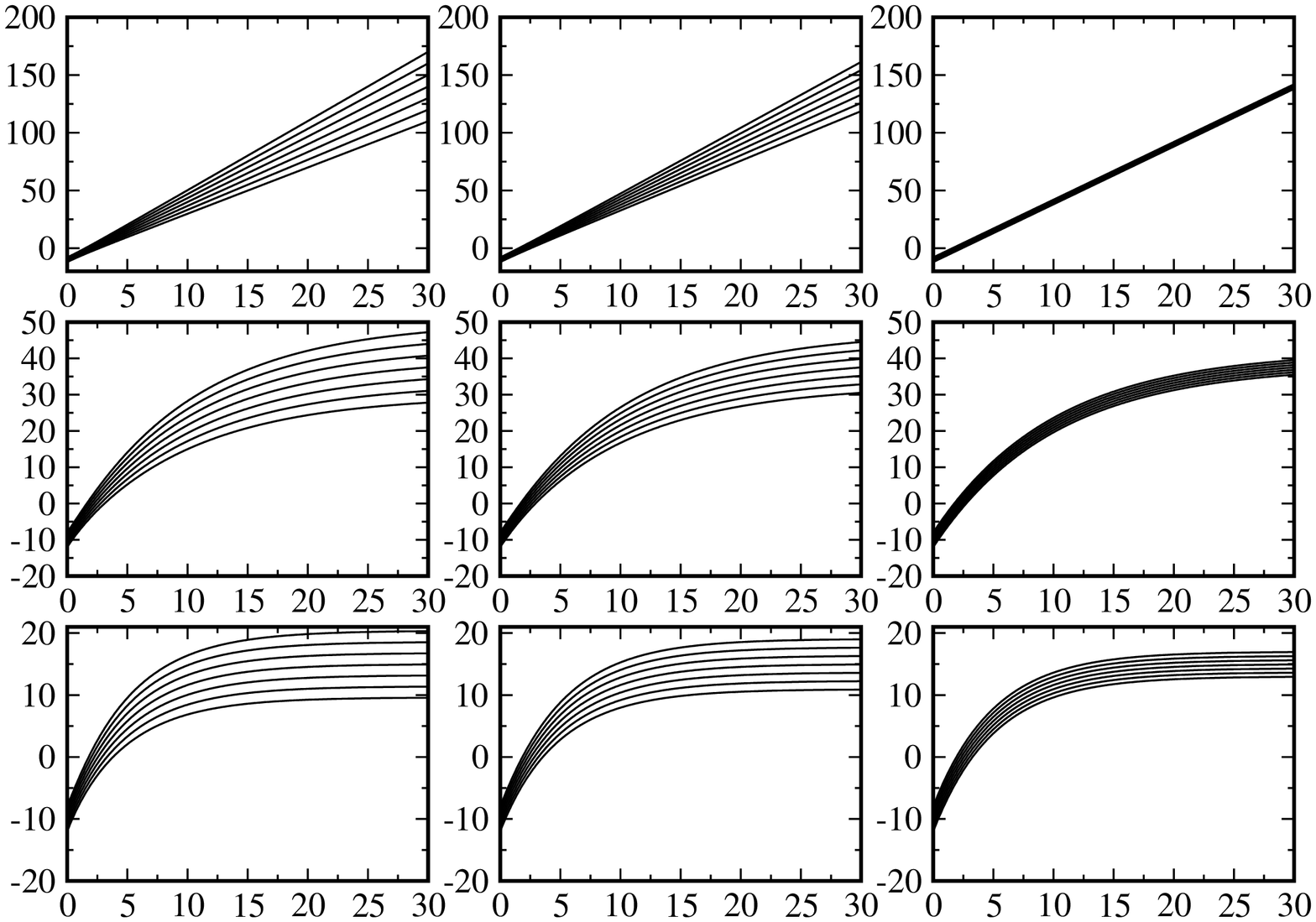}
\caption{
Scaled trajectories  issued from a freely propagating  Gaussian wave packet in a viscous medium. 
In each row, the friction coefficient $\gamma$ is 
constant: $\gamma = 0$  (first row), $\gamma = 0.1$ (second row) and $\gamma = 0.2$ (third row).
In each column, the dyanmical regime $\epsilon$ is constant:  $\epsilon = 1$ (first column), 
$\epsilon = 0.5$ (second column), $\epsilon = 0$ (third column).}
\label{fig: trajectories_free}
\end{figure}

\begin{figure}[h]
\centering
\includegraphics[width=10cm,angle=-90]{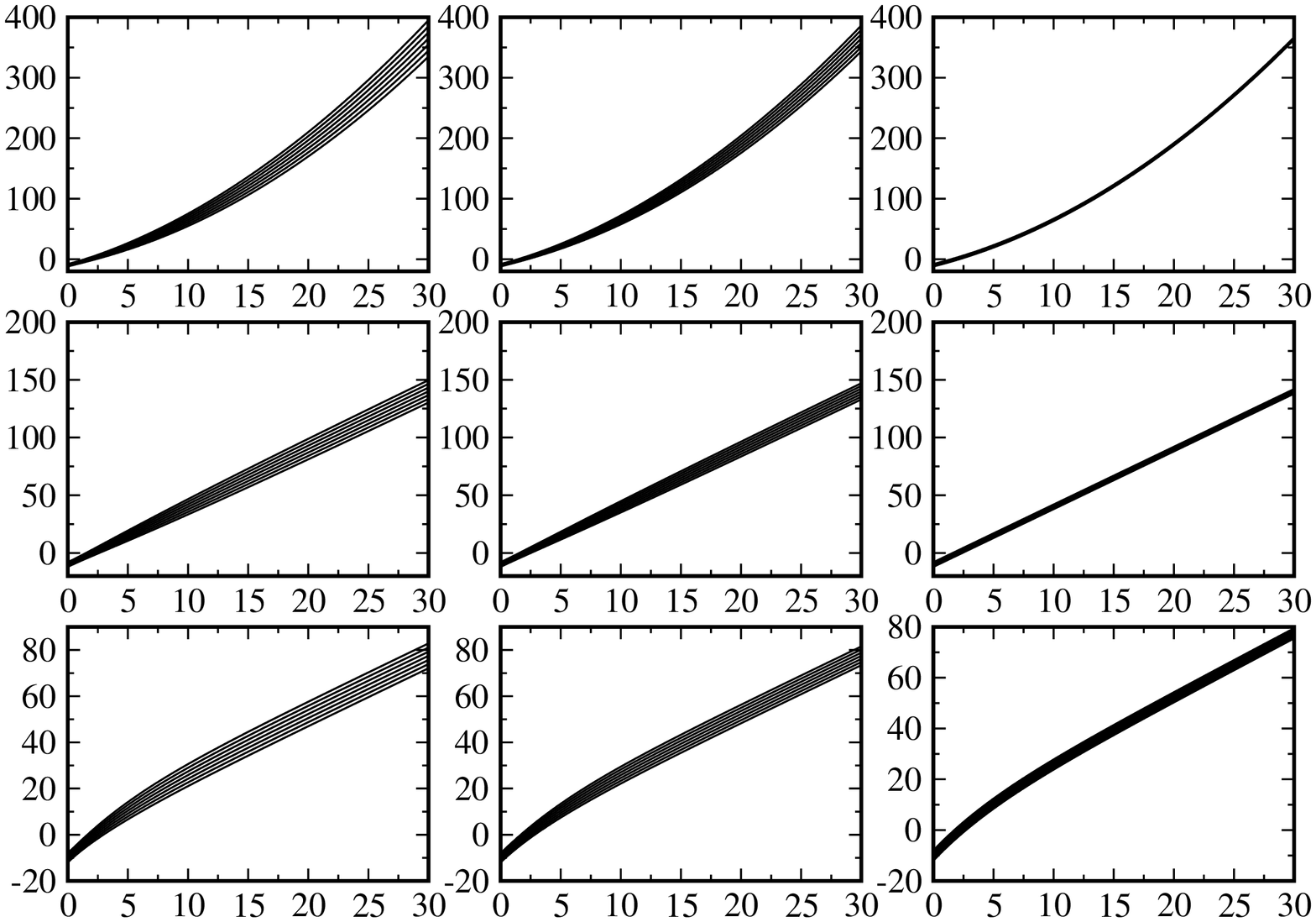}
\caption{
Scaled trajectories issued from a propagating  Gaussian packet in a viscous medium
under the presence of a constant force filed ($a = -0.5$). 
In each row, the friction coefficient $\gamma$ is 
constant: $\gamma = 0$  (first row), $\gamma = 0.1$ (second row) and $\gamma = 0.2$ (third row).
In each column, the dyanmical regime $\epsilon$ is constant:  $\epsilon = 1$ (first column), 
$\epsilon = 0.5$ (second column), $\epsilon = 0$ (third column).}
\label{fig: trajectories_linear}
\end{figure}

In the following two figures, Figs. \ref{fig: trajectories_free} and \ref{fig: trajectories_linear}, the 
corresponding scaled trajectories derived from Eqs. (\ref{eq: BM traj_free_viscid}) 
and (\ref{Bohm_traj_linear}) are plotted. In each row, the friction coefficient $\gamma$ is 
constant: $\gamma = 0$  (first row), $\gamma = 0.1$ (second row) and $\gamma = 0.2$ (third row).
In each column, the dyanmical regime $\epsilon$ is constant:  $\epsilon = 1$ (first column), 
$\epsilon = 0.5$ (second column), $\epsilon = 0$ (third column).
%
%
%
Scaled trajectories do not cross and bend with friction as a function of time.  
Furthermore, the distance between two specific Bohmian trajectories, from Eqs. 
(\ref{eq: BM traj_free_viscid}) and (\ref{Bohm_traj_linear}) can be expressed as
\begin{eqnarray} \label{eq: traj_distance}
x(x_1^{(0)}, t) - x(x_2^{(0)}, t) &=& \left( x_1^{(0)} - x_2^{(0)} \right) \frac{\ti{\sigma}_t}{\sigma_0}
\end{eqnarray}
for the free and linear potential cases.
Since $\ti{\sigma}_t$ increases with time then trajectories diverge in the quantum case, but 
width of the wave packet in the classical case is constant meaning that in this case the distance 
between trajectories remains constant; trajectories never cross each other even in the classical 
case. This feature is apparent in Figures \ref{fig: trajectories_free} and \ref{fig: trajectories_linear}.
It should also be noticed that for a given time, $\ti{\sigma}_t$ decreases with $\epsilon$ and
the rate of divergence of trajectories decreases continuously along the quantum-classical  transition.

The corresponding Bohmian and classical trajectories are identical  to those of Ref. 
\cite{SaCaRoLoMi-AP-2014} obtained from a given anzat for the wave function.

\subsection{Propagation of a scaled Gaussian wave packet in a damped harmonic potential}

Let us analyze now the time evolution of the Gaussian wave packet (\ref{eq: inital_Gauss}) in 
the damped harmonic potential $ \hat{H} = \dfrac{\hat{p}^2}{2m} 
e^{-\gamma t} + \dfrac{1}{2} m \omega_0^2 \hat{x}^2 e^{\gamma t} $. 
By using Eqs. (\ref{eq: wavet_propag_wave0}) and (\ref{eq: harmonic_CK_propagator}), 
we obtain
\begin{eqnarray} \label{eq: wavet_har}
\ti{\psi}(x, t) &=& \frac{1}{ (2\pi  \tilde{s}_t^2)^{1/4} }
\exp \left[ \alpha_t (x-x_t)^2 + \frac{i}{\tilde{\hbar}} ~p_t ~e^{\gamma t} (x-x_t) + \frac{i}{\tilde{\hbar}} ~ \eta_t \right]   ,
\end{eqnarray}
where
\begin{eqnarray} \label{eq: st_har}
\tilde{s}_t &=&  \sigma_0 ~ e^{-\gamma t /2} \left( \cos \omega t + \frac{\gamma}{2} \frac{\sin\omega t}{\omega}
+ i \frac{\ti{\hbar}}{2m\sigma_0^2} \frac{\sin\omega t}{\omega} \right) ,
\end{eqnarray}
and
\begin{eqnarray*}
\alpha_t &=& i \frac{m}{2\ti{\hbar}} e^{\gamma t} \left[ -\frac{\gamma}{2}
 + \frac{\omega}{\sin \omega t} \left( \cos \omega t - \frac{\sigma_0}{\ti{s}_t} e^{-\gamma t /2} \right) \right] , \\
\eta_t &=& \frac{m}{4} \frac{\sin^2\omega t}{\omega^2}
\left[ 2 ~ \omega \cot(\omega t) \left( \frac{p_0^2}{m^2} - \omega_0^2 x_0^2 \right)
- \gamma \left( \frac{p_0}{m} + \frac{\gamma x_0}{2} \right)^2 - 4 \omega^2 x_0 \left( \frac{p_0}{m} + \frac{ \gamma x_0}{4} \right)
\right] ,
\end{eqnarray*}

with $\omega$, $x_t$ and $p_t$ given respectively by
\begin{eqnarray}
\omega &=& \sqrt{\omega_0^2 - \frac{\gamma^2}{4}} , \label{eq: omega_omega0&gamma} \\
x_t &=& e^{-\gamma t/2}
\left[  x_0 \left( \cos \omega t + \frac{\gamma}{2} \frac{\sin \omega t}{\omega} \right)
+ \frac{p_0}{m} \frac{\sin \omega t}{\omega} \right] , \label{eq: xt_har} \\
p_t &=& m \dot{x}_t = e^{-\gamma t/2}
\left[  - m\omega_0^2 x_0 \frac{\sin \omega t}{\omega} + p_0\left( \cos \omega t - \frac{\gamma}{2} \frac{\sin \omega t}{\omega} \right) \right] , \label{eq: pt_har}
\end{eqnarray}
which corresponds to the classical trajectory $x_t$ for the classical damped harmonic oscillator 
with initial position $x_0$ and initial momentum $p_0$. Only the underdamped case, 
$\omega_0 > \gamma / 2$, is considered here. The center of the Gaussian wave 
packet follows $x_t$ with  velocity $p_t/m$. One can easily check that $ \langle \hat{x} \rangle(t) 
= x_t$ and $ \langle \hat{p} \rangle(t)~ e^{-\gamma t} = p_t$ which are acceptable results 
according to the Ehrenfest theorem. One can also see that
\begin{eqnarray} \label{eq: alpha_rest}
\alpha_0 &=& - \frac{1}{4 \sigma_0^2} ,
\end{eqnarray}
which is a reasonable result due to the relation $ \ti{\psi}(x, 0) =\ti{\psi}_0(x) $.

%
\begin{figure}[h]
\centering
\includegraphics[width=10cm,angle=-90]{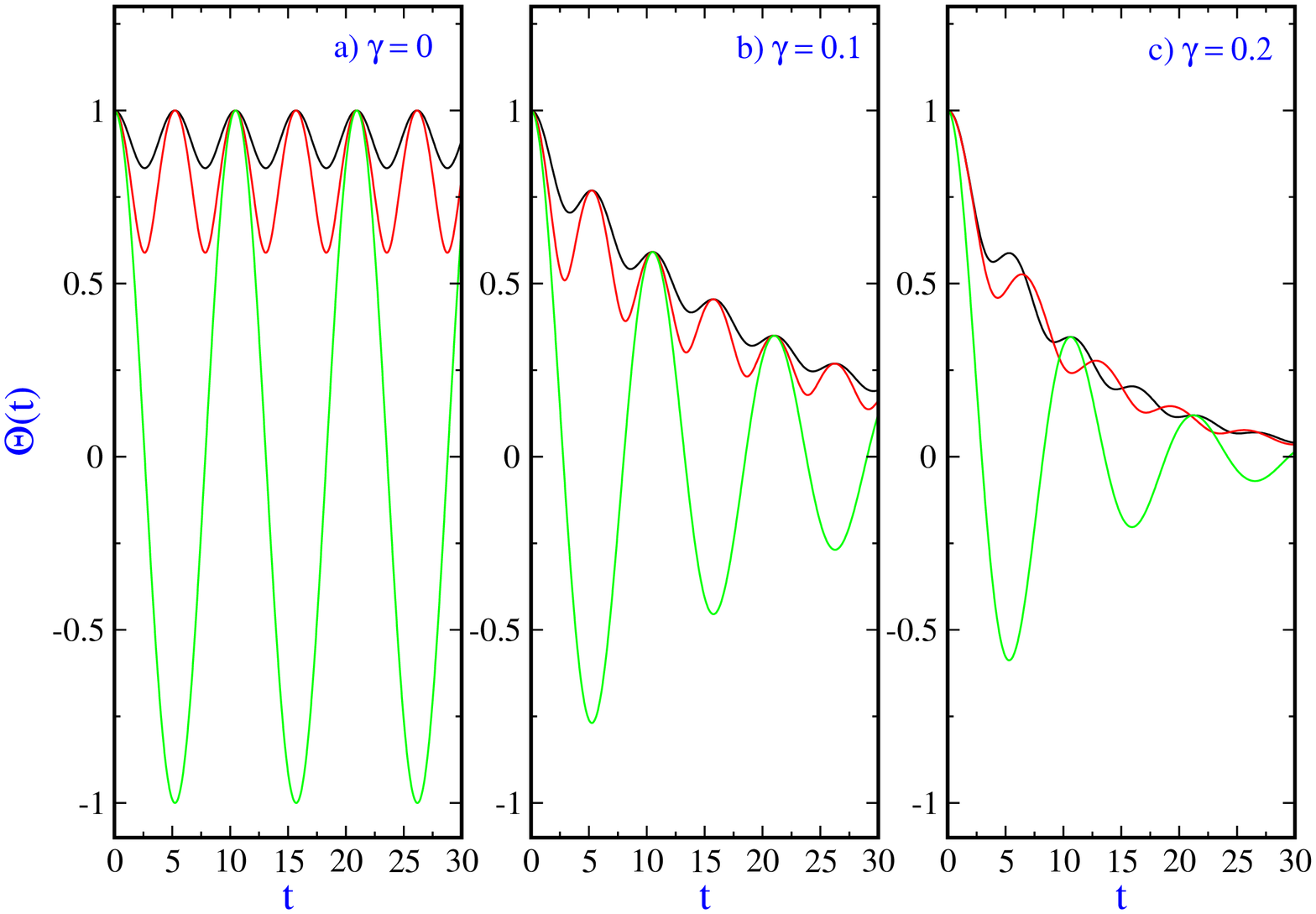}
\caption{(Color online)
$\Theta (t)$ is plotted as a function of time for different values of $\epsilon$ and  $\gamma$ for 
$\omega_0 = 0.6$.  The different dynamical regimes are: $\epsilon = 1$ (Bohmian, black curve), 
$\epsilon = 0.5$ (intermediate, red curve), and $\epsilon = 0$ (classical, green curve).}
\label{fig: Theta_func}
\end{figure}
%
\begin{figure}[h]
\centering
\includegraphics[width=10cm,angle=-90]{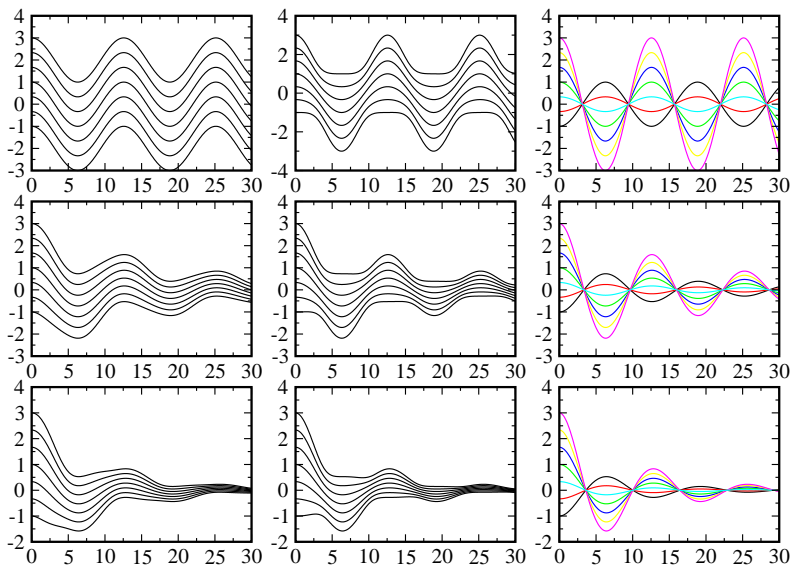}
\caption{(Color online)
Scaled trajectories issued from a  propagating  Gaussian packet in a viscous medium
under the presence of a harmonic potential ($\omega_0 = 0.5$). In each row 
dissipation factor $\gamma$ is constant while in each column transition parameter $\epsilon$ 
is constant: $\gamma = 0$ (first row), $\gamma = 0.1$ (second row), $\gamma = 0.2$ (third row);  
$\epsilon = 1$ (first column), $\epsilon = 0.5$ (second column), $\epsilon = 0$ (third column).}
\label{fig: trajectories_harmonic}
\end{figure}

The probability density is given by Eq. (\ref{eq: rho_general}) with $x_t$ given by 
Eq. (\ref{eq: xt_har}) and width 
\begin{eqnarray} \label{eq: sigmat_har}\
\ti{\sigma}_t &=& \sigma_0~ e^{-\gamma t /2} \sqrt{ \left( \cos \omega t + \frac{\gamma}{2} \frac{\sin\omega t}{\omega}  \right)^2 + \frac{\epsilon \hbar^2}{4 m^2 \sigma_0^4 } \frac{\sin^2\omega t}{\omega^2} } .
\end{eqnarray}
Finally, the scaled velocity field is given by
\begin{eqnarray}  \label{eq: vB_har}
v_s(x, t) &=& \theta(t)~ (x-x_t) + \frac{p_t}{m} ,
\end{eqnarray}
from which the corresponding trajectories are given by
\begin{eqnarray} \label{Bohm_traj_har}
x(x^{(0)}, t) &=& x_t + (x^{(0)} - x_0) ~ \Theta(t) ,
\end{eqnarray}
where
\begin{eqnarray} \label{eq: theta_function}
\theta(t) &=& \frac{2 \ti{\hbar}}{m} ~ \text{Im}( \alpha_{t} ) ~ e^{-\gamma t} =
 -\frac{\gamma}{2} + \omega \cot \omega t - \frac{\sigma_0^2}{\ti{\sigma}_t^2} ~e^{-\gamma t}
\left(  \frac{\gamma}{2} + \omega \cot \omega t \right), \\
\Theta(t) &=& \exp \left[ \int_0^t dt' ~ \theta(t') \right] .
\end{eqnarray}
Here, the dressing scheme is again reproduced but in a more involved way through  $\Theta (t)$. 
The physical meaning of this function is better understood from the frictionless and classical limits.
The density function for the actual momentum has also the Gaussian shape 
(\ref{eq: mom_dis__generalform}) but with $p_t$ given by Eq. (\ref{eq: pt_har}) and the width 
by
\begin{eqnarray} \label{eq: Pi_p_har}
\ti{\Sigma}_t &=& m \sigma_0 | \dot{\Theta} (t)| = m \sigma_0 | \theta(t)  ~ \Theta(t) | .
\end{eqnarray}
Note that even for $ \omega_0 \leq \frac{\gamma}{2} $ where $ \omega $ is zero or imaginary, 
still $\ti{\sigma}_t$, $\theta(t)$,  $\Theta(t)$ and $ \ti{\Sigma}_t $ are all real functions of time.
From Eq. (\ref{eq: alpha_rest}), it follows that $\ti{\Sigma}_0 = 0 $, i.e., the actual momentum 
distribution function is initially  a Dirac delta function $\Pi(p, 0) = \delta(p-p_0)$, meaning that all 
particles of the ensemble have the same momentum $p_0$.


For propagation in a non-viscous medium, $\gamma = 0$, one obtains
\begin{eqnarray} \label{eqs: non-viscid_Har}
\ti{\sigma}_t &=& \sigma_0~ \sqrt{
\cos^2 \omega_0 t  +  \frac{\epsilon \hbar^2}{4 m^2 \sigma_0^4 } \frac{\sin^2\omega_0 t}{\omega_0 ^2} } , \\
\Theta(t) &=& \frac{ \ti{\sigma}_t }{\sigma_0} ,\\
x(x^{(0)}, t) &=& x_t + (x^{(0)} - x_0) ~ \frac{ \ti{\sigma}_t }{\sigma_0} ,\\
\tilde{\Sigma}_t &=& m \dot{\ti{\sigma}}_t .
\end{eqnarray}
where $\Theta (t)$ gives the ratio of the Gaussian wave packet width at $t$ and $t=0$. This 
function is precisely the ratio appearing in Eq. (\ref{Bohm_traj_generalform}) as a factor 
of the second sum of the dressing scheme, which can be seen as a normalized time dependent width.
In the classical limit $ \epsilon \rightarrow 0 $ we have
\begin{eqnarray}
\tilde{\sigma}_t &=&  \sigma_0 ~ e^{-\gamma t /2} \left| \cos \omega t + \frac{\gamma}{2} \frac{\sin\omega t}{\omega}
\right| , \\
\theta(t) &=& -\frac{ \omega_0^2 }{ \omega \cot \omega t + \gamma/2} ,\\
\Theta(t) &=& e^{-\gamma t /2} \left( \cos \omega t + \frac{\gamma}{2} \frac{\sin\omega t}{\omega} \right) , \label{eq: Theta_cl} \\
\ti{\Sigma}_t &=& m \sigma_0 \omega_0^2 \left| \frac{\sin\omega t}{\omega} \right| ~ e^{-\gamma t /2}
= m \frac{\omega_0^2 }{ \left|  \frac{\gamma}{2} + \omega \cot \omega t \right| } \tilde{\sigma}_t .
\end{eqnarray}
where again $\Theta (t)$ is given by the normalized time dependent width. In this case, this
width displays a damped oscillatory motion with the frequency given by $\omega$. 
Thus, we can claim that $\Theta (t)$, for the scaled
trajectories, represents the time evolution of the normalized width of the Gaussian wave packet, which
is a purely wave property.  
Moreover, by considering  Eq. (\ref{eq: Theta_cl}) in (\ref{Bohm_traj_har}), the 
dissipative classical trajectory is expressed as
\begin{eqnarray} \label{eq: cl_traj_damped har}
x_{\text{cl}}(x^{(0)}, t) &=& x_t + (x^{(0)} - x_0)
e^{-\gamma t /2} \left( \cos \omega t + \frac{\gamma}{2} \frac{\sin\omega t}{\omega} \right) ,
\end{eqnarray}
indicating that all particles of the classical ensemble have initially the same momentum.
According Eq. (\ref{Bohm_traj_har}), $\Theta (t)$ could also be seen as the 
deviation (or difference) of the scaled trajectory with respect to the $x_t$-trajectory.
It is noticeable to observe that widths do not reduce to constant values in the classical limit.
In Figure \ref{fig: Theta_func}, $\Theta (t)$ is plotted as a function of time for different values 
of $\epsilon$ and  $\gamma$ and for $\omega_0 = 0.6$.  The different dynamical regimes are: 
$\epsilon = 1$ (black curve), $\epsilon = 0.5$ (red curve), and $\epsilon = 0$ (green curve). It 
is clearly seen the oscillatory behaviour displayed by this difference. For the classical regime,
this difference becomes zero since the trajectories can cross. However, 
the scaled trajectories do not cross because $\Theta (t)$ approaches asymptotically to 
zero but never reaches it.
As a result of this behaviour, trajectories are becoming more dense at larger friction coefficients. 
These features are clearly observed in Figure \ref{fig: trajectories_harmonic}.
In this case, we have taken a motionless wave packet, $ p_0 = 0 $, with center at $ x_0 = 1 $ for 
numerical calculations. The corresponding scaled trajectories are plotted for the same values of
$\gamma$ and $\epsilon$ as before.
%
%
The resulting trajectories are also identical to those of Ref. \cite{SaCaRoLoMi-AP-2014} 
obtained from a given anzat for the wave function when $\epsilon =0,1$.
\section{Concluding remarks}

The field of open quantum systems is being very active and fruitful to better undertand 
the decoherence process. In this work, we have used one of the three approaches menioned 
in the Introduction, the effective Hamiltonian approach, where all the degrees of freedom
of the bath coordinates have been suppressed. The CK model represents the paradigm for 
dissipative dynamics since it is simple and very intuitive. When it is applied to simple interaction
potentials, the dynamics is also solved by analytical methods.  For this reason,  we 
have used this model to analyze the quantum-classical transition through trajectories in a 
continuous way. Starting from the nonlinear classical wave equation for conservative systems 
due to Rozen and introducing a scaled Planck's contant, the dissipative dynamics is proposed 
to be described  by a linear Schr\"odinger equation. This differential equation is solved following
the Bohmian procedure in order to obtain what we call scaled trajectories. The continuous parameter
$\epsilon$ responsible for the smooth quantum-classical transition covers the interval $[0,1]$, where
$\epsilon = 1$ provides the standard Bohmian (quantum) trajectories and $\epsilon =0$, the 
corresponding classical ones. These scaled trajectories follows always the same dressing scheme
consisting of a classical trajectory (particle property) plus a term involving the width of the wave
packet (wave property). Thus, the decoherence as well as localization can be seen as gradual 
processes enabling us the understanding and visualization of this dissipative dynamics. This procedure
could be easily extended to deal with other interesting problems such as, for example,
tunneling, gas collisions at high presure and surface diffusion.


\appendix

\section{}
In this appendix some usefull relations are given within the CK model.

\subsection{Classical wave equation} \label{app: class_wave}

By using the polar form of classical wave function, $ \psi_{\text{cl}}(x, t) = R_{\text{cl}}(x, t) 
\exp [ i S_{\text{cl}}(x, t)/\hbar ] $ in the classical wave equation (\ref{eq: class_wave equation}), 
we obtain two coupled differential equations
\begin{eqnarray}
\frac{\partial  R_{\cl}}{\partial t} &=&  -\frac{1}{2m } \left( 2 \frac{\partial  R_{\cl}}{\partial x} \frac{\partial  S_{\cl}}{\partial x} + R_{\cl} \frac{\partial^2  S_{\cl}}{\partial x^2} \right) e^{-\gamma t} ,
\\
- \frac{\partial  S_{\cl}}{\partial t} &=& \frac{1}{2m} \left( \frac{\partial  S_{\cl}}{\partial x} \right)^2 e^{-\gamma t} + V(x) e^{\gamma t} . \label{eq: class_HJ}
\end{eqnarray}
By multiplying both sides of the first equation by $2 R_{\text{cl}}$ and re-arranging terms, 
one obtains the continuity equation
\begin{eqnarray} \label{eq: class_continuity}
\frac{\partial  R^2_{\cl}}{\partial t} + \frac{\partial}{\partial x}\left( R^2_{\cl} \frac{1}{m} \frac{\partial  S_{\cl}}{\partial x} e^{-\gamma t} \right)  &=& 0 ,
\end{eqnarray}
whereas Eq. (\ref{eq: class_HJ}) is the classical Hamilton-Jackobi equation. Noting the 
continuity equation (\ref{eq: class_continuity}), the classical velocity is defined as
\begin{eqnarray} \label{eq: class_velocity}
\dot{x} &=& \frac{1}{m} \frac{\partial  S_{\cl}}{\partial x} e^{-\gamma t} ,
\end{eqnarray}
from which one obtains the classical equation of motion
\begin{eqnarray} \label{eq: class_eq_motion}
m\ddot{x} &=& \frac{d}{dt} \left( \frac{\partial  S_{\cl}}{\partial x} e^{-\gamma t} \right)
=  e^{-\gamma t} \left[ -\gamma \frac{\partial S_{\cl} }{\partial x} +
\left( \frac{\partial  }{\partial t} + \dot{x} \frac{\partial  }{\partial x} \right) \frac{\partial S_{\cl} }{\partial x}
\right]
\nonumber \\
&=&
e^{-\gamma t} \left[ -\gamma \frac{\partial S_{\cl} }{\partial x}  + \frac{\partial}{\partial x} \frac{\partial S_{\cl} }{\partial t} + \frac{1}{m} \frac{\partial  S_{\cl}}{\partial x} e^{-\gamma t} \frac{\partial^2  S^2_{\cl}}{\partial x}
\right]
\nonumber \\
&=&
e^{-\gamma t} \left[ - m \gamma \dot{x} e^{\gamma t}
- \frac{\partial}{\partial x} \left( \frac{1}{2m} \left( \frac{\partial  S_{\cl}}{\partial x} \right)^2 e^{-\gamma t} + V(x) e^{\gamma t} \right) + \frac{1}{m} \frac{\partial  S_{\cl}}{\partial x} e^{-\gamma t} \frac{\partial^2  S^2_{\cl}}{\partial x}
\right]
\nonumber \\
&=&
- m \gamma \dot{x} - \frac{\partial V}{\partial x} ,
\end{eqnarray}
where we have used Eqs. (\ref{eq: class_velocity}) and the partial derivative of 
(\ref{eq: class_HJ}) with respect to the space coordinate.


\subsection{Propagators}
\label{app: propag}

Since the Hamiltonian of the CK model in free space commutes with itself in different times, 
the formal solution of the Schr\"{o}dinger equation is given by \citep{Sakurai-book-2011}
\begin{eqnarray} \label{eq: wavet_propag_wave0}
\psi(x, t) &=& \int dx' G(x, x'; t) \psi(x', 0) ,
\end{eqnarray}
where
\begin{eqnarray} \label{eq: propagator_formula}
G(x, x'; t) &=& \langle x | e^{-i\int_0^t dt' H(t')/\hbar} | x' \rangle
\end{eqnarray}
is called the propagator of the system. For the free Hamiltonian of CK model, one has
\begin{eqnarray*}
G_{\text{free}}(x, x'; t) &=&  \langle x |
\exp \left[ -\frac{i}{\hbar}\int_0^t dt' \frac{\hat{p}^2}{2m} e^{-\gamma t'} \right]
| x' \rangle
\\
&=&
 \langle x |
\exp \left[ -\frac{i}{\hbar}\int_0^t dt' \frac{\hat{p}^2}{2m} e^{-\gamma t'} \right]
\int_{-\infty}^{\infty} dp' |p'\rangle \langle p'
| x' \rangle
\\
&=& \frac{1}{ 2\pi \hbar } \int_{-\infty}^{\infty} dp' e^{i p' (x-x') / \hbar } \exp \left[ -\frac{i}{\hbar} \frac{p'^2}{2m} \int_0^t dt' e^{-\gamma t'} \right] ,
\end{eqnarray*}
where we have used the fact that
\begin{eqnarray*}
\langle x | p \rangle &=& \frac{1}{ \sqrt{2\pi \hbar} } e^{ i p x/\hbar} ,
\end{eqnarray*}
%
Finally, by calculating the  integrals, one obtains
\begin{eqnarray} \label{eq: free_CK_propagator}
G_{\text{free}}(x, x'; t) &=& \sqrt{ \frac{m\gamma}{2\pi i \hbar (1-e^{-\gamma t})} }
\exp \left[ \frac{i m}{2\hbar} \frac{\gamma}{1-e^{-\gamma t}} (x-x')^2 \right]
\end{eqnarray}
for freely propagating particle in viscid media within the CKframework. In the non-viscid limit, 
$\gamma \rightarrow 0$, (\ref{eq: free_CK_propagator}) gives the known result of the free 
particle propagator. It should be mentioned that in the case of scaled Schr\"{o}dinger equation 
one must replace $\hbar$ by $\ti{\hbar}$ in the above equations.

It is remarkable that in the linear potential, Hamiltonian $ \hat{H}(t) = \frac{\hat{p}^2}{2m} 
e^{-\gamma t} + m a ~ \hat{x} e^{\gamma t}$ does not commute at different times, 
$ [\hat{H}(t_1), \hat{H}(t_2)] = 2 i \hbar a~ \sinh[\gamma(t_1-t_2)] ~ \hat{p} $. Thus, 
the evolution operator $U(t)$ does not have the simple form $\exp\left[ -\frac{i}{\hbar} 
\int_0^t dt' \hat{H}(t') \right]$. Therefore, by the above method we cannot compute the propagator 
of the linear potential in CK model.

Moeeira \cite{Mo-LNC-1978} has already computed the propagator of linear potential 
$V(x) = m a ~ \hat{x}$ within the framework of CK using the Lagrangian formulation:
\begin{eqnarray}
G(x, x'; t) &=& \exp\left[ -\frac{i m}{ \hbar} a \left( x ~ \frac{e^{\gamma t}-\gamma t -1}{\gamma(1-e^{-\gamma t})}
+ x' ~ \frac{e^{-\gamma t}+\gamma t -1}{\gamma(1-e^{-\gamma t})} \right)
 -\frac{i m}{2\hbar} a^2 ~ \frac{e^{\gamma t}+e^{-\gamma t}-\gamma^2 t^2-2}{\gamma^3(1-e^{-\gamma t})}
\right]
\nonumber \\
&\times&  G_{\text{free}}(x, x'; t)
\label{eq: linear_CK_propagator}
\end{eqnarray}
which has the correct form in the non-viscid limit.

Moeeira \cite{Mo-LNC-1978} has also given propagator for the damped harmonic oscillator 
with Hamiltonian
$ \hat{H} = \dfrac{\hat{p}^2}{2m} e^{-\gamma t} + \dfrac{1}{2} m \omega_0^2 \hat{x}^2 e^{\gamma t} $
as follows
\begin{eqnarray}
G(x, x'; t) &=&  \sqrt{ \frac{ m \omega e^{\gamma t/2} }{ 2\pi i \hbar \sin(\omega t)} }
\exp\left[ \frac{i m}{ 4 \hbar} \gamma \left( x'^2 - x^2 e^{\gamma t} \right)
+ \frac{ i m \omega }{ 2\hbar\sin(\omega t) } [ ( x^2 e^{\gamma t} + x'^2 ) \cos(\omega t) - 2 x x' e^{\gamma t/2} ]
\right]
\nonumber \\
\label{eq: harmonic_CK_propagator}
\end{eqnarray}
with $ \omega = \sqrt{ \omega_0^2 - \gamma^2/4 }$.


\subsection{Fourier transform of the scaled free Gaussian wave packet} \label{ap: four_trans}

As usual, we define the Fourier transform of the scaled wave function $\ti{\psi}(x, t)$ as follows,
\begin{eqnarray} \label{eq: Fourier_trans}
\ti{\phi}(p, t) &=& \frac{1}{ \sqrt{2 \pi \ti{\hbar} } } \int dx ~ e^{-i p x / \ti{\hbar}} \ti{\psi}(x, t).
\end{eqnarray}

At first, we derive the scaled Schr\"{o}dinger equation in momentum space. To this end, we 
compute the partial derivative of $\ti{\phi}(p, t)$ with respect to time
\begin{eqnarray*}
i \ti{\hbar} \frac{\partial}{\partial t} \ti{\phi}(p, t) &=&
\frac{1}{ \sqrt{2 \pi \ti{\hbar} } } \int dx ~ e^{-i p x / \ti{\hbar}}
\left( i \ti{\hbar} \frac{\partial}{\partial t} \ti{\psi}(x, t) \right) \\
&=&
\frac{p^2}{2m} e^{-\gamma t}  \ti{\phi}(p, t) +
\frac{1}{ \sqrt{2 \pi \ti{\hbar} } } \int dx ~ e^{-i p x / \ti{\hbar}} e^{\gamma t} V(x)
\ti{\psi}(x, t),
\end{eqnarray*}
where we have used the scaled Schr\"{o}dinger equation (\ref{eq: Scaled Sch}) and the 
integration by parts. Thus, for the free particle in viscid media, the  wave function in momentum 
space satisfies
\begin{eqnarray} \label{eq: Sch_free_eq_mom}
i \ti{\hbar} \frac{\partial}{\partial t} \ti{\phi}(p, t) &=&
\frac{p^2}{2m} e^{-\gamma t}  \ti{\phi}(p, t) .
\end{eqnarray}

Now, we compute the Fourier transform of the free damped scaled Gaussian wave packet 
(\ref{eq: scaled psi_G(t)}) using (\ref{eq: Fourier_trans})
\begin{eqnarray} \label{eq: Fourier_trans_free_Gauss}
\ti{\phi}(p, t) &=& \frac{1}{ \sqrt{2 \pi \ti{\hbar} } } \int dx ~ e^{-i p x / \ti{\hbar}}
\frac{1}{(2 \pi \ti{s}_t^2)^{1/4}}
\exp \left[ - \frac{(x-x_t)^2}{4\sigma_0 \ti{s}_t} + \frac{i}{\ti{\hbar}} p_0( x - x_t ) + \frac{i}{ \ti{\hbar} } \frac{p_0^2}{2m}  \frac{1-e^{-\gamma t}}{\gamma} \right]
\nonumber \\
&=&
\frac{1}{(2 \pi \ti{\sigma_p}^2)^{1/4}}
\exp \left[ - \frac{(p-p_0)^2}{ 4\ti{\sigma_p}^2 } - \frac{i}{\ti{\hbar}} p ~ x_0 - \frac{i}{ \ti{\hbar} } \frac{p^2}{2m}  \frac{1-e^{-\gamma t}}{\gamma} \right] ,
\end{eqnarray}
with
\begin{eqnarray} \label{eq: mom_width_mom_wf}
\ti{\sigma_p} &=& \frac{\ti{\hbar}}{2 \sigma_0} .
\end{eqnarray}
Thus, the canonical momentum distribution function reads
\begin{eqnarray} \label{eq: can_mom_dis}
| \ti{\phi}(p, t) |^2 &=&  \frac{1}{ \sqrt{2 \pi} \ti{\sigma_p} }
\exp \left[ - \frac{(p-p_0)^2}{2 \ti{\sigma_p}^2 } \right] .
\end{eqnarray}
This distribution function has a time-independent width $ \ti{\sigma}_p $ and is centered at 
$p_0$ and is independent on the friction coefficient $\gamma$.
%

\vspace{2cm}
\noindent
{\bf Acknowledgements}
SVM acknowledges partial support from the University of Qom and SMA support from 
the Ministerio de Econom\'ia y Competitividad (Spain) under the Project 
FIS2014-52172-C2-1-P. 

%
%
%

%
\end{document}